# Extended Gauge Theory, Bi-Spinors and Scalar Supersymmetry

A. Jourjine[1]

FG CTP
Hofmann Str. 6-8
01281 Dresden, Germany

Abstract

Within the context of the extended bi-spinor gauge theory we describe a new off-shell realization of scalar supersymmetry (s-susy) of massless interacting fields with $U(1)$, $U(1) \times SU(N)$, and $U(1) \times SU(N_1) \times SU(N_2)$ gauge groups. S-susy acts in the space of graded differential forms. The realization is non-linear in the non-abelian case. S-susy would not require the doubling of the SM particle spectrum. The theory is by construction globally $U(2,2) \approx U(1) \times SO(2,4)$ invariant and is an example of a supersymmetric conformal field theory.

Supersymmetry offers a promising way to look at a number of the Standard Model (SM) problems: the hierarchy problem, the gauge coupling unification, the Dark Matter origin, the inclusion of gravity. In all present supersymmetric extensions of the SM a particle must have an observable superpartner with a spin differing by one half, which results in the on-shell doubling of the observed particle spectrum. Despite an intensive search, none of the super-particles have been found [1].

In this letter we develop further an alternative approach to supersymmetry where the extra to the SM field content required for supersymmetry is largely not observable: the propagating bosonic sector contains only Higgs fields and gauge fields plus a scalar which can couple only to gravity. The not propagating extra filed content in the fermionic sector is the $4^{th}$ generation of quarks and leptons.

In [2], [3], we described free scalar supersymmetry (free s-susy) for non-interacting fields described by supersymmetric bi-spinor gauge theories. These lie in the intersection of seemingly unrelated areas of the extended gauge theories and the bi-spinor field theory. The former appear often in superstring and supergravity theories [4, 5], the latter first were used to describe the electron spin [6], then interpret it geometrically [7], [8], and in the lattice fermion doubling problem [9, 10].

Free s-susy uses a bi-spinor description of fermions. The description is very restrictive and leads to observable relations between the elements of the flavor mixing matrices for quarks and leptons. At tree level it predicts the TBM form of the PMNS matrix and the $V_{ts} = V_{cb}, V_{cs} = V_{tb}$ relations for the CKM matrix, which are actually observed within the experimental errors. The three generation unitarity is violated due

---

to the fourth generation but this violation is within the experimental error for the current unitarity tests.

The basic bosonic and fermionic fields in bi-spinor gauge theory are described by differential forms. The bosonic sector is described by an extended gauge theory with a minimal free differential algebra, in other words, by differential forms with values in the Lie algebra of the gauge group, whereas the fermionic sector by four generations of fermions. Free s-susy is broken by interactions. Unlike for the standard supersymmetry, the supersymmetric charges of s-susy are Lorenz scalars, hence its name. In a certain basis for differential forms, the z-basis, equivalence between the standard susy and free s-susy can be established. However, if the spin structure on the space-time cannot be defined then the equivalence is only local. On-shell the field content of free s-susy additional to the SM is essentially the fourth generation of fermions. One puzzling feature of free s-susy is that it manifests itself only in the Lorenz gauge. In this Letter we describe a different realization of massless s-susy that does not have this defect. In addition we show in examples of $U(1)$ and $U(1) \times SU(N)$ that it can be defined for interacting fields. An intriguing feature of non-abelian s-susy is that the $U(1)$ factor is required and that for the non-abelian case the spinor content has to be chiral, which is indeed observed experimentally, while Majorana spinors are ruled out. Also intriguing is the fact that the theory has a global $U(2,2)$ invariance. Since $U(2,2) \approx SO(2,4)$ the theory is in effect an example of supersymmetric CFT.

Unless specifically mentioned, we use the conventions of [2, 3] and work with four-dimensional manifolds. Extension of our results to other space-time manifolds with dimensions other then four is straightforward. For Euclidean space-times $U(2,2)$ should be replaced by $U(4)$. An introduction to the bi-spinor QFT can be found in [11]. Derivation of the constraints on the elements of the flavor mixing matrices for quarks and leptons can be found in [12].

To treat fermions as differential forms we need to describe the z-basis in the space of differential forms. In the standard coordinate basis (c-basis) a differential form is represented as $\Phi = \sum \Phi_{[\mu_p]} dx^{[\mu_p]}$, $[\mu_p] = \mu_1, \cdots, \mu_p$, $\mu_1 < \cdots < \mu_p$. The $z$ - basis is defined by the equation $(d - \delta)Z = iZ\gamma^\mu \partial_\mu$ on the matrix of differential forms $Z$, where $d, -\delta$ are the exterior derivative and its adjoint with respect to the scalar product

$$\langle \Phi, \Phi' \rangle = \int d^4x \sqrt{-\det g} \, tr \left[ \sum_p \alpha \, \Phi^+_{[\mu_p]} \Phi'^{[\mu_p]} \right], \qquad (1)$$

where trace is taken over the Lorenz and, where appropriate, over the gauge indices. Operator $\alpha : \alpha \Phi_p = (-1)^p \Phi_p$ is absent in the Euclidean case. One finds that $Z = \sum_p \gamma_{[\mu_p]} dx^{[\mu_p]}$ [10].

Using the two bases we can write $\Phi = \sum \Phi_{[\mu_p]} dx^{[\mu_p]}$ or $\Phi = tr[Z \Psi(\Phi)]$. The relation between the coefficients is $\Psi = (1/4)\sum_p \Phi_{[\mu_p]} \gamma^{[\mu_p]}$, $\Phi_{[\mu_p]} = tr[\gamma_{[\mu_p]} \Psi]$. Note that $\Phi_{[\mu_p]}$ are antisymmetric tensor fields, while $\Psi_{\alpha\beta}(\Phi)$ is a bi-spinor. As a result, in the bi-spinor gauge theory off-shell bi-spinors and gauge fields can be distinguished only by their commutativity properties.

To establish the connection with the standard quantum filed theory we need to describe how to extract the spinor degrees of freedom from a bi-spinor $\Psi$. This can be done by the spinbein decomposition. For $U(1)$ the decomposition is defined using a spinbein $\eta^A$, $A = 1,\ldots,4$, a multiplet of normalized commuting classical Dirac spinors [13]. The spinbein decomposition of $\Psi$ into its spinor components $\psi^A, \eta^A$ is the *anzatz*

$$\Psi_{\alpha\beta}(x) = \psi_\alpha^A(x)\overline{\overline{\eta}}_\beta^A(x), \tag{2}$$

where $\eta^A$ is normalized according to

$$\overline{\overline{\eta}}_\alpha^A \eta_\alpha^B = \delta^{AB}, \tag{3}$$

$$\overline{\overline{\eta}}^{aA} = \Gamma^{AB}\overline{\eta}^{aB},\ \Gamma^{AB} = \Gamma_{AB} = diag(1,\ 1,-1,-1) \in U(2,2). \tag{4}$$

For $U(1)$ spinbein $\eta^A(x)$ is invertible and can be chosen to be constant. Spinbein decomposition with a constant spinbein transfers globally all physical degrees of freedom from a bi-spinor to a flavor multiplet of spinors, as can be seen from $tr\overline{\overline{\Psi}}(i\partial + gA)\Psi = \overline{\overline{\psi}}^A(i\partial + gA)\psi^A$, where $\overline{\overline{\Psi}}$ the bi-spinor conjugate of $\Psi$, $\overline{\overline{\Psi}} = \gamma^0 \Psi^+ \gamma^0$. It turns out that two generations of $\psi^A$ propagate forward in time, while the remaining two propagate backward in time [11].

For the non-abelian case spinbein decomposition is more elaborate. Spinbeins are no longer invertible nor can they be chosen to be constant globally in a gauge-invariant way. However, the form of the spinbein *ansatz* is the same. Since $\psi^A, \eta^A$ transform in the $N$ of the $SU(N)$ in $U(N) = U(1) \times SU(N)$ and $\Psi$ transforms in $N \times \overline{N}$ we now define

$$\Psi_{\alpha\beta}^{ab}(x) = \psi_\alpha^{aA}(x)\overline{\overline{\eta}}_\beta^{Ab}(x), \qquad \overline{\overline{\eta}}_\alpha^{Aa}\eta_\alpha^{aB} = \delta^{AB}, \tag{5}$$

where $a, b$ are the gauge indices.

Note that the right hand side of (5) is invariant under

$$\psi^A(x), \eta^A(x) \to U^{AB}(x)\psi^B(x),\ U^{AB}(x)\eta^B(x),\ U \in U(2,2).$$

Therefore, the choice of spinbein to a degree is a matter of convenience. Using this freedom of choice we can make equations of motion for each $\psi^A$ to be exactly the same as for a spinor. To see this consider bi-spinor action $S = \int d^4x\, tr\overline{\overline{\Psi}}(i\partial + gA_1)\Psi$ with the equation of motion $i\partial\Psi + g[A_1, \Psi] = 0$. Comparing this with $i\partial\hat{\psi}^A + gA_1\hat{\psi}^A = 0$ for Dirac spinors $\hat{\psi}^A$, we see the difference in the presence of the commutator. However, after spinbein *ansatz* (5) we obtain new bi-spinor equations of motion

$$(i\partial + gA_1)\psi^A - \gamma^\mu \psi^B\left[-i(\overline{\overline{\eta}}^B \partial_\mu \eta^A) + g(\overline{\overline{\eta}}^B A_\mu \eta^A)\right] = 0. \tag{6}$$

We can now choose a spinbein so that the second term in (6) vanishes. It is sufficient to take $\eta^A$ proportional to the holonomy for the connection $A_1 = A_\mu dx^\mu$

$$\eta^A(x) = T_C \exp\left(-ig \int_{C_x} A_\mu(y) dy^\mu\right) \eta_0^A, \quad \bar{\eta}_0^{Aa} \eta_0^{aB} = \delta^{AB}, \tag{7}$$

where $\eta_0^A = const$ and $T_C$ is the path ordered product along some loop $C_x$ punctured at $x$, i.e., it has an infinitesimal gap at $x$. As required $\eta^A(x)$ constructed according to (7) is a spinor which transforms in the fundamental representation of the gauge group. We conclude that setting a spinbein by (7) we can regard $\psi^A$ as a multiplet of four spinors that obey Dirac equations of motion. At the same time (7) breaks the local $U(2,2)$ invariance to its global subgroup. It is this global $U(2,2)$ that leads to the appearance of additional scalar spin quantum number for elementary fermions, which in turn leads to the approximately observed relations between the elements of CKM and PMNS mixing matrices. The combination of (5) and (7) defines the spinbein decomposition for the non-abelian case. Note that our argument applies to the gravitational interaction as well so long as they are described as a gauge theory.

Having described how to extract the physical spinor degrees of freedom from bi-spinors, we proceed with s-susy. In the extended bi-spinor gauge theory gauge fields are commuting differential forms $A = \sum_{p=0}^{p=4} A_p$, $A_p = A_{[\mu_p]} dx^{[\mu_p]}$ with values in some Lie algebra. Formally, gauge potentials $A_p$ form a minimal free differential algebra [5]. Curvature of $A$ is defined as a form $F$, $F = \sum_{p=0}^{p=3} F_{p+1}$, where $F_{p+1}$ is the curvature of gauge $p$-form $A_p$. It is usually assumed that the expansion of $F$ starts with $p=1$. In the supersymmetric bi-spinor gauge theory we have to include the $p=0$ term for s-susy to work. Only then the fermionic and the bosonic degrees of freedom can be matched.

For $U(1)$ the $p$-curvatures $F_p$ are defined by

$$F_{p+1} = dA_p, \tag{8}$$

while for $SU(N)$

$$F_1 = dA_0, \tag{9}$$

$$F_2 = d - i g A_1 \wedge A_1, \tag{10}$$

$$F_{p+1} = dA_p - i g [A_1 \wedge, A_p], \quad p > 1. \tag{11}$$

where $g$ is the coupling constant, $\wedge$ is the exterior product, and the commutator is the Lie algebra commutator. $F_2$ is the usual curvature 2-form of the 1-form connection $A_1$.

In the non-abelian case, as a consequence of $d^2 = 0$ curvatures $F_p$ satisfy the Bianchi identities. In our case they are

$$dF_1 = 0, \tag{12}$$

$$d_{A_1} F_2 = 0, \tag{13}$$

$$d_{A_1} F_3 - ig[A_2, F_2] = 0, \tag{14}$$

$$dF_4 = 0, \tag{15}$$

where the gauged exterior derivative $d_{A_1}$ is defined by

$$d_{A_1} = d - ig[A_1 \wedge, \cdot] \tag{16}$$

except when it acts on $A_1$ when the commutator is absent.

In the abelian case the Bianchi identities (12-15) simplify to $dF_p = 0$.

In the SM the gauge fields transform in the adjoint representation of $SU(N)$, or $U(1) \times SU(N_1) \times SU(N_2)$, while fermions transform in the fundamental representations and hence bi-spinors in $N \times \overline{N}$, where $N \times \overline{N} = 1 \oplus (N^2 - 1)$, and $(N_1 \cdot \overline{N}_1) \times (N_2 \cdot \overline{N}_2) = 1 \oplus (N_1^2 - 1) \oplus (N_2^2 - 1) \oplus ((N_1^2 - 1) \times (N_2^2 - 1))$ in the irreducible components. These identities give a hint about how to build irreducible s-susy multiplets. We see at once that a $U(1)$ factor is required for s-susy. Only then the number of degrees of freedom for the bosonic and the fermionic fields can be matched. Further, if the bosonic gauge fields are hermitean a match is possible only if fermions carry exactly the half of the degrees of freedom of the complex fundamental representation.

In order to write down the s-susy in detail it is convenient to decompose the bosonic and the fermionic forms $A, \Phi$ into traceful and traceless components. For $U(1) \times SU(N)$ we define

$$A^{ab} = \delta^{ab} B + W^{ab}, \quad tr W = 0, \tag{17}$$

$$\Phi^{ab} = \delta^{ab} tr \Phi + \hat{\Phi}^{a,b}, \quad tr \hat{\Phi} = 0, \tag{18}$$

where $B$ is the $U(1)$ gauge field and $W$ is the $SU(N)$ gauge field. For $U(1) \times SU(N_1) \times SU(N_2)$ we can define

$$A^{ai,bj} = \delta^{ab} \delta^{ij} B + W^{ab} \delta^{ij} + \delta^{ab} G^{ij} + \hat{A}^{ai,bj}, \tag{19}$$

$$\Phi^{ai,bj} = \delta^{ab} \delta^{ij} tr_{1,2} \Phi + (tr_2 \Phi)^{ab} \delta^{ij} + \delta^{ab} (tr_1 \Phi)^{ij} + \hat{\Phi}^{ai,bj}, \tag{20}$$

$$tr_1 W = 0, \quad tr_2 G = 0, \quad tr_1 \hat{A} = tr_2 \hat{A} = 0, \quad tr_1 \hat{\Phi} = tr_2 \hat{\Phi} = 0, \tag{21}$$

where $B$, $W$, $G$ are gauge fields and $\hat{A}$, $\hat{\Phi}$ are the traceless fields. The index of the trace denotes over which Lie algebra the trace is taken.

We now have all the ingredients to describe s-susy of the bi-spinor extended gauge theory for $U(1)$ and $U(1) \times SU(N)$. As can be seen from (19) irreducible decomposition in the case $U(1) \times SU(N_1) \times SU(N_2)$ contains $\hat{A}^{ai,bj}$, a gauge field for which there is no experimental interpretation. Therefore, s-susy variation $\delta_\varepsilon$ for $U(1) \times SU(N_1) \times SU(N_2)$ must be realized by putting $\hat{A}^{ai,bj} = 0$, and setting $\delta_\varepsilon \hat{A}^{ai,bj} = 0$, $\delta_\varepsilon \hat{\Phi}^{ai,bj} = 0$. Since decompositions (19-21) are invariant under all symmetries of the theory, such constraints are also invariant and thus are consistent.

The action of the $U(1) \times SU(N)$ theory describes the dynamics of the bosonic commuting differential form $A$ and the fermionic anti-commuting differential form $\Phi$. It is given by

$$S = S_b + S_f = c_1 \langle F, F \rangle + c_2 \langle \Phi, (d_{A_1} - \delta_{A_1}) \Phi \rangle, \quad F = F_B + F_W, \qquad (22)$$

where $\langle .,. \rangle$ is defined by (1) and $c_{1,2}$ are numerical coefficients, which for now we leave unspecified.

Action (22) contains action for the gauge field 1-form $A_1 = A_\mu dx^\mu$ interacting in the gauge invariant way with a multiplet of four spinors. The Lorentz gauge-fixing condition is $\partial_\mu A^\mu \equiv -\delta_{A_1} A_1 = 0$, where $-\delta_{A_1}$ is the adjoint of $d_{A_1}$ with respect to (1). It is natural to generalize it to

$$\delta_{A_1} A = 0. \qquad (23)$$

Action (22) is manifestly gauge invariant with respect to the infinitesimal extended gauge transformations parameterized by a differential form $\lambda = \sum_{p=0}^{4} \lambda_p$

$$\delta A_0 = const, \qquad \delta F_1 = 0, \qquad (24)$$

$$\delta A_1 = d_{A_1} \lambda_0, \qquad \delta F_2 = [F_2, \lambda_0], \qquad (25)$$

$$\delta A_p = d_{A_1} \lambda_{p-1} + [A_p, \lambda_0], \quad \delta F_p = [F_p, \lambda_0] + [F_2, \lambda_{p-2}], \quad p = 3,4. (26)$$

The $p = 0$ case gives the standard gauge transformations of $A_1$ and its curvature $F_2$. In addition $\lambda$ contains terms with $p > 0$ which are unrelated to the fermion gauge transformations.

Further, component $A_0$ is a Lorentz scalar on which the gauge-fixing condition $\delta_{A_1} A = 0$ places no restrictions, because $\delta_{A_1} A_0 \equiv 0$. This component is decoupled from the rest of the theory, because $\langle F_1, F_1 \rangle = \partial_\mu A_0 \partial^\mu A_0$. However, the equivalence principle requires that it is coupled to gravity.

Component $A_4$ is absent from the first term in (22) because $d_{A_1} A_4 \equiv 0$. Therefore, $A_4$ is an auxiliary field. Its elimination introduces a potential function into the action. Note that, in principle, we can generalize (16) to $d_{A_1} = d - igA_1 - ig'A_4$. This is the only possible generalization that is consistent with gauge symmetry. Such a generalization will lead to additional potential terms after elimination of $A_4$.

It is a common knowledge that gauge theory of abelian $p$-forms in $D$ dimensions is equivalent to a theory of $D - p - 2$ forms [14]. In our case this means that for $U(1)$ the $A_3$ field is non-dynamical, while $A_2$ field is equivalent to $A_0$ field. However, unlike $A_0$ the field $A_2$ is coupled to $A_1$ in a gauge-invariant way. In the end in the bosonic sector we are left with the standard gauge field $A_1$ plus a scalar $A_0$ that couples only to gravity and field $A_2$ that is coupled to $A_1$ and is a Lorenz scalar. Similar analysis for the non-abelian case is lacking, however it is plausible. Therefore, interpreted in the context of the SM and cosmology, the extra content of our theory contains a prototype for a Higgs field and, depending on whether $A_0$ is massive or not, a prototype of either dark energy or dark matter.

After these preliminary remarks we can state our main result. Namely, that using (12-15) it is straightforward to prove that (22) is invariant with respect to s-susy transformations that mix bosonic and fermionic fields. For the $U(1)$ and the $U(1) \times SU(N)$ cases s-susy is described as follows. Given two anticommuting Lorentz scalar parameters $\varepsilon_{1,2}$ the infinitesimal s-susy transformation is

$$\delta_\varepsilon A = \varepsilon_1 \Phi, \qquad \delta_\varepsilon \Phi = \varepsilon_2 F, \tag{27}$$

where for $U(1) \times SU(N)$ one uses the decomposition (17, 18) and $F = F_B + F_W$. For hermitean gauge fields action (22) is invariant under (27) provided that

$$c_1 \varepsilon_1^* + c_2 \varepsilon_2 = 0, \qquad \varepsilon_1^* + \varepsilon_1 = 0. \tag{28}$$

As for fermion fields, for the $U(1)$ case we have three possibilities

(1) $\Phi_p^+ = \Phi_p$, \hfill (29)

(2) $\Phi_0 = \Phi_2 = \Phi_4 = 0$, $\Phi_1, \Phi_3$ - arbitrary complex, \hfill (30)

(3) $\Phi_1 = \Phi_3 = 0$ \quad $\Phi_0 = \Phi_2 = \Phi_4$ - arbitrary complex. \hfill (31)

Each of the three possible conditions cuts the number of fermionic degrees of freedom in half. It is easy to see that with appropriate definition of the basis in the space of gamma matrix algebra these possibilities correspond to Majorana, left chiral and right chiral bi-spinors. Hence, we call the corresponding supermultiplets the Majorana, the left and the right chiral supermultiplets, respectively. For the $U(1) \times SU(N)$ case the first possibility is ruled out and there are only the left and the right chiral supermultiplets. Thus, if it can be used for extension of the SM, s-susy predicts chirality of the elementary fermions. Majorana spinors would then be allowed only for the inert neutrinos.

For the abelian case $d_{A_1}$ reduces to $d$ and in the Lorenz gauge one can establish a one-to-one correspondence of (27) with the standard susy [13]. The constraint (29) relates to the c-basis. In the z-basis from $\Psi = (1/4)\sum_p \Phi_{[\mu_p]} \gamma^{[\mu_p]}$, $\Phi_{[\mu_p]} = tr[\gamma_{[\mu_p]} \Psi]$ and $\gamma$-matrix properties we obtain the constraints on the bi-spinor field $\Psi$

$$\overline{\overline{\Psi}} = (1/4)\sum_p (-1)^{\beta_p} \Phi_{[\mu_p]} \gamma^{[\mu_p]}, \qquad \left(\overline{\overline{\Psi}}\right)^T \equiv C\Psi \qquad (32)$$

where $\overline{\overline{\Psi}} = \gamma^0 \Psi^+ \gamma^0$, $\beta_p = (1/2)p(p-1)$ and $C$ is the matrix defined by the first equality.

Using (5, 7) from (32) we obtain constrains on the four physical Dirac spinor fields $\psi^A = \Psi \eta^A$ determined using (2, 3) or (5). We obtain from (29) that

$$\overline{\psi}^A = \overline{\eta}^A \overline{\overline{\Psi}} = \overline{\eta}^A \left(\psi^B \overline{\eta}^B\right)^T C^T \equiv \left(C\psi^A\right)^T. \qquad (33)$$

In the standard QFT the analogous linear relation between a spinor and its Dirac conjugate uniquely defines a Lorenz-invariant subspace of Majorana spinors. Therefore, relation (29) constrains $\psi^A$ to be Majorana spinors. At the same time the (anti)commutativity of $\gamma^5$ with (odd)even-numbered products of $\gamma$-matrices implies that bi-spinors constructed with constraint (30) are of definite chirality, while those under constraint (31) are of opposite chirality. This property is inherited by the Dirac spinors derived from bi-spinors. We conclude that (30, 31) describe chiral or Weyl spinors.

For $U(1) \times SU(N_1) \times SU(N_2)$ we note that in this case we need to use decomposition (19-21) and $F = F_B + F_W + F_G$. Then one can construct s-susy corresponding to the two subgroups $U(1) \times SU(N_1)$ and $U(1) \times SU(N_2)$.

We conclude with a brief discussion of possible applications of scalar supersymmetry. Clearly, to describe an extension of the SM action (22) has to be much better understood and possibly modified. If an s-susy extension for the SM can be constructed, one would need to confirm the desirable properties of supersymmetry in its scalar version, including viable scenarios for its breaking. It is not clear that the potential function that is already present in (22) leads to mass generation. Another big s-susy question is where is the fourth generation? In its defense one can say that the extra fourth generation seems somehow more plausible then the double of the already observed particle spectrum, because it is a more economical description of supersymmetry. Another intriguing detail comes to attention when one considers the fact that the four-generational s-susy has two generations that propagate forward in time and two generations that propagate backward in time. As a result the theory is manifestly time reflection invariant. Absence of the fourth generation means that time invariance is broken. Since CPT = 1 in any reasonable QFT, this in turn may be interpreted as CP violations. Therefore, within the context of s-susy QFT the absence of the fourth generation and the observed CP violation are a direct consequence of one another. In any case, in the least s-susy could offer a fresh look at the old problems with the standard supersymmetry.

# References

[1] C. Autermann, Experimental Status of Supersymmetry after the LHC Run-I, arXiv:1609.01686.

[2] A. Jourjine, Scalar Supersymmetry, Physics Letters B 727 (2013) 211, arXiv:1208.1867.

[3] A. Jourjine, Scalar Supersymmetry in Bi-spinor Gauge Theory, arXiv:1302.6162.

[4] P. Salgado and S. Salgado, Extended gauge theory and gauged Free Differential Algebras, arXiv:1702.07819.

[5] in L. Castellani, R. D'Auria and P. Fré, "Supergravity and Superstring: a geometric perspective", World Scientific, Singapore 1991.

[6] D. Ivanenko and L. Landau, (1928), Zeitschrift f. Physik, 48, 340.

[7] I. M. Benn and R. W. Tucker, Phys. Lett. B117 (1982) 348.

[8] T. Banks, Y. Dothan, D. Horn, Phys. Lett. B 117 (1982) 413.

[9] H. B. Nielsen, M. Ninomija, Phys. Lett. B 105, 219 (1981), Nucl. Phys. B 185, 20 (1981).

[10] P. Becher and H. Joos, Z. Phys. C15 (1982) 343-365.

[11] A. Jourjine, The Quantum Theory of Bi-Spinor Fields, arXiv:1306.17.

[12] A. Jourjine, Scalar Spin of Elementary Fermions, Phys. Lett. B728 (2014) 347, arXiv:1307.2694.

[13] A. Jourjine, Mass mixing, the fourth generation, and the kinematic Higgs mechanism, Phys. Lett. B 693 (2010) 149, arXiv:1005.3593.

[14] S. Weinberg, The Quantum Theory of Fields I, Ch. 8.8, Cambridge University Press, 2005.